\pdfoutput=1 
\documentclass[aps,prl,twocolumn,superscriptaddress,amsfont,graphicx,nofootinbib,preprintnumbers]{revtex4}
\usepackage{amsmath}
\usepackage{color}
\usepackage{graphicx}
\usepackage{hyperref}
\usepackage{blindtext}
\usepackage{slashed}

\definecolor{nicered}{rgb}{0.7,0.1,0.1}
\definecolor{nicegreen}{rgb}{0.0,0.4,0.0}
\hypersetup{colorlinks,citecolor=nicegreen,linkcolor=nicered,urlcolor=nicered}

\newcommand{\h}{\boldsymbol{\lambda}}

\newcommand{\rd}{\mathrm d}

\newcommand{\LC}{\mathrm{LC}}

\newcommand{\calA}{\mathcal{A}}
\newcommand{\calB}{\mathcal{B}}

\newcommand{\calR}{\mathcal{R}}

\newcommand{\eps}{\epsilon}

\newcommand{\lb}{\left(}
\newcommand{\rb}{\right)}

\newcommand{\nfvo}{n^{\gamma}_f}
\newcommand{\nfvt}{n^{\gamma\gamma}_f}

\begin{document}

\def\OX{Rudolf Peierls Centre for Theoretical Physics, University of Oxford,
Clarendon Laboratory, Parks Road, Oxford OX1 3PU}
\def\MSU{Department of Physics and Astronomy, Michigan State University,
East Lansing, Michigan 48824, USA}

\preprint{MSUHEP-21-010, OUTP-21-12P}

\title{Two-loop helicity amplitudes for diphoton plus jet production in full color}

\author{Bakul Agarwal}            
\email[Electronic address: ]{agarwalb@msu.edu}
\affiliation{\MSU}

\author{Federico Buccioni}            
\email[Electronic address: ]{federico.buccioni@physics.ox.ac.uk}
\affiliation{\OX}

\author{Andreas von Manteuffel}            
\email[Electronic address: ]{vmante@msu.edu}
\affiliation{\MSU}

\author{Lorenzo Tancredi}            
\email[Electronic address: ]{lorenzo.tancredi@physics.ox.ac.uk}
\affiliation{\OX}

\begin{abstract}
We present the complete two-loop corrections in massless QCD
for the production of two photons and a jet, taking into account all color structures.
In particular, we analytically compute all two-loop helicity amplitudes for the quark-antiquark, quark-gluon, and antiquark-gluon channel, and check them with an 
independent calculation of the polarization-summed interference with the tree amplitude.
This is the first time that 
two-loop QCD corrections to a five-point scattering process have
been computed beyond the leading-color approximation for all helicity configurations.
\end{abstract}

\maketitle

The past two decades have witnessed impressive developments in our understanding of the properties of 
scattering amplitudes in quantum field theory, both in supersymmetric theories such as $\mathcal{N}=4$ Super Yang Mills,
and in phenomenologically relevant ones such as Quantum Chromodynamics (QCD).  
The increasing interest devoted by
physicists and mathematicians to the study of the formal properties of scattering amplitudes 
is justified by their role in providing us with essential building blocks  
to interpret 
the experimental data produced at particle colliders like the CERN Large Hadron Collider (LHC).
The experiments running at LHC are being extremely successful at measuring  
a multitude of interesting physical observables with unprecedented precision, opening the way to
a new era of precision collider physics. In particular, comparing high-quality data with precise theoretical
predictions for wisely constructed physical observables offers a promising opportunity to unveil
so-far elusive signs of new physics beyond the Standard Model.
The required 
theoretical calculations can be conveniently performed in perturbative quantum field theory by 
means of series expansions in the relevant coupling constants, which are assumed to be small. 
The coefficients of the perturbative series are in turn expressed in terms of Feynman diagrams 
of increasing complexity, i.e.\ with increasing numbers of external legs and internal loops.

While a general mathematical understanding of scattering amplitudes in perturbative quantum field theory
remains beyond our reach, important breakthroughs have been achieved in the past years, allowing us to 
tame their mathematical complexity.
In particular, thanks to the interplay of techniques borrowed from particle physics phenomenology~\cite{Tkachov:1981wb,Chetyrkin:1981qh,Kotikov:1990kg,Bern:1993kr,Remiddi:1997ny,Gehrmann:1999as,Papadopoulos:2014lla} and $\mathcal{N}=4$ Super Yang Mills~\cite{Dixon:1996wi,ArkaniHamed:2010gh,Kotikov:2010gf,Henn:2013pwa} 
together with developments in the theory of special functions in quantum field theory~\cite{Goncharov,Remiddi:1999ew,Goncharov:2001iea,Goncharov:2010jf,Brown:2008um,Ablinger:2013cf,Panzer:2014caa,Duhr:2011zq,Duhr:2012fh,Duhr:2019tlz}
which originated from branches of pure mathematics such as
number theory and algebraic geometry, we now possess a toolkit that 
allows us to compute 
most relevant scattering amplitudes for $2 \to 2$ processes up to two loops in QCD. 
The impressive experimental 
measurements that are being performed by the LHC experiments
have contributed to pushing the theory calculations further, 
moving the frontier to two-loop corrections for $2 \to 3$ processes and three-loop corrections for  
$2 \to 2$ processes in QCD.
Indeed, on one hand, the first calculations of three-loop 
corrections to $2 \to 2$ massless
processes in supersymmetric theories~\cite{Henn:2020lye,Henn:2016jdu,Henn:2019rgj} and, very recently, in full QCD~\cite{Caola:2020dfu} have been completed. 
On the other hand, considerable effort has been put in understanding the properties of 
two-loop $2 \to 3$ processes,
which went hand-in-hand with important breakthroughs both in the development of new techniques for handling their algebraic complexity~\cite{Hodges:2009hk,Ita:2015tya,Badger:2016uuq,Abreu:2017xsl,Chawdhry:2018awn,Abreu:2020xvt,vonManteuffel:2014ixa,Peraro:2016wsq,Peraro:2019svx},
and in the study of the special functions required for their 
calculation~\cite{Papadopoulos:2015jft,Gehrmann:2018yef,Chicherin:2018mue,Chicherin:2020oor}.

Thanks to these advancements, many results have been obtained in recent years. In particular, 
many publications have been devoted to the calculation of so-called leading-color corrections to scattering amplitudes for various $2 \to 3$
processes~\cite{Gehrmann:2015bfy,Badger:2017jhb,Abreu:2017hqn,Abreu:2018aqd,Abreu:2018zmy,Abreu:2018jgq,Abreu:2019rpt,Abreu:2020cwb,Badger:2018enw,Chicherin:2018yne,Chicherin:2019xeg,Chawdhry:2020for,DeLaurentis:2020qle,Agarwal:2021grm,Badger:2021nhg,Abreu:2021fuk,Chawdhry:2021mkw}.
The leading-color approximation often provides a realistic estimate of the bulk of the corrections 
while avoiding 
the complications arising from
non-planar two-loop diagrams.
In turn, these amplitudes have made it possible to perform  the first phenomenological studies in next-to-next-to-leading-order (NNLO) leading-color QCD for the production of three photons at the LHC~\cite{Chawdhry:2019bji,Kallweit:2020gcp}.
Although all non-planar master integrals for $2 \to 3$
massless processes are now
available~\cite{Chicherin:2020oor}, 
their use in the actual computation of 
physical scattering amplitudes has remained, as of today, an outstanding task.
Specifically, the algebraic complexity grows dramatically for non-planar diagrams and, 
as a consequence, full results in QCD have been obtained only for five-gluon scattering in the substantially simpler
all-equal helicity configuration~\cite{Badger:2019djh}. 
The latter, however, does not possess all the features of a generic two-loop 
five-point process.

In this Letter, we fill the gap and compute, for the first time, 
the full-color two-loop QCD corrections for a $2 \to 3$ scattering process at the LHC.
We consider, in particular, all helicity amplitudes for the production of two photons and a 
strongly interacting parton in parton-parton scattering.
This class of processes is of particular interest phenomenologically: confronting measurements of pairs of photons with non-zero transverse momentum at the LHC with precise theory predictions makes it possible to put stringent constraints on various beyond-standard-model scenarios. 
Furthermore, from a technical point of view, 
the production of two photons and a parton is highly non-trivial,
specifically due to the presence of colored particles in both the initial and final states. 
Finally, the two-loop corrections considered in this Letter
constitute the last missing amplitudes for the calculation of the N$^3$LO QCD corrections 
to the production of two photons at the LHC.

The production of two photons and a parton in proton-proton collisions can proceed through three 
classes of partonic channels, namely quark-antiquark annihilation, $q\bar{q}\to g \gamma \gamma$,
quark(antiquark)-gluon scattering, $q(\bar{q}) g \to q(\bar{q}) \gamma \gamma$, and the loop-induced
gluon-fusion channel, $g g \to g \gamma \gamma$.
The two-loop corrections to the latter formally contributes to 
the N$^3$LO and N$^4$LO cross sections for di-photon plus jet and di-photon production respectively, 
thus we do not consider them in this Letter.
We focus instead on the channels that involve a pair
of quarks. 
In order to have a notation for the kinematics that is as uniform as possible, 
we define the
three sub-processes as
\begin{align}
\label{eq:qqbgaaprocess}
q(p_1) + \bar{q}(p_2) & \to g(p_3) + \gamma(p_4) + \gamma(p_5)\,, \\
\label{eq:qgqaaprocess}
q(p_1) + g(p_2)  & \to q(p_3) + \gamma(p_4) + \gamma(p_5)\,, \\
\label{eq:gqbaaprocess}
g(p_1) + \bar{q}(p_2)  & \to \bar{q}(p_3) + \gamma(p_4) + \gamma(p_5)\,,
\end{align}
where we stress the different assignment 
of momenta for the partons in the three channels.
All external particles are massless, $p^2_i=0$ for $i=1,\ldots,5$, thus the kinematics is 
completely fixed in terms 
of five independent kinematic invariants, which we choose as
\begin{gather} 
  s_{12}=(p_1+p_2)^2, \quad  s_{23}=(p_2-p_3)^2, \quad s_{34}=(p_3+p_4)^2, \nonumber \\
  s_{45}=(p_4+p_5)^2, \quad  s_{15}=(p_1-p_5)^2 \,.
  \label{eq:indinvs}
\end{gather}
For all processes, the physical scattering region is identified by
$s_{12},s_{34},s_{45} >0$ and $s_{23},s_{15} < 0$,
together with the conditions on the invariants~\cite{Gehrmann:2018yef}
\begin{gather}
s_{12} \geq s_{34}\,, \quad s_{12} - s_{34} \geq s_{45}\,, \nonumber \\ 
 0 \geq s_{23} \geq s_{45}-s_{12}\,, \quad s_{15}^- \leq s_{15} \leq s_{15}^+\,,
\end{gather}
where
\begin{align*}
s_{15}^\pm &= \frac{1}{(s_{12}-s_{45})^2} \Big[ s_{12}^2 s_{23} + s_{34} s_{45} (s_{45} - s_{23}) 
\nonumber \\
&
- s_{12} (s_{34} s_{45} + s_{23} s_{34} + s_{23} s_{45})
\nonumber \\
&\pm \sqrt{s_{12} s_{23} s_{34} s_{45} (s_{12} + s_{23} - s_{45})(s_{34} + s_{45} - s_{12})}\, 
\Big].
\end{align*}
In order to fully describe the helicity amplitudes for these processes, it 
is also useful to introduce the parity-odd invariant
\begin{equation}
  \label{eq:epsfive}
  \epsilon_5 = 4 i\, \epsilon_{\mu\nu\rho\sigma} p^\mu_1 p^\nu_2 p^\rho_3 p^\sigma_4\,,
\end{equation}
where $\epsilon_{\mu\nu\rho\sigma}$ is the totally anti-symmetric Levi-Civita symbol,
$(\epsilon_5)^2 =  \Delta$, and
\begin{align}
\label{eq:Gram}
    \Delta &= - 4 s_{12} s_{23} s_{34} (s_{23} - s_{45} - s_{15}) \nonumber \\ 
    &+ (s_{12} s_{23} + s_{23} s_{34} - s_{34} s_{45} + s_{45} s_{15} - s_{15} s_{12})^2\,,
\end{align}
is the determinant of the Gram matrix $(G_{ij}) = (2 p_i p_j)$.
In the scattering physical region, one finds $\Delta<0$
and $\epsilon_5 = \pm i \sqrt{|\Delta |}$, 
where the sign depends on the actual kinematics inside this region~\cite{Byers:1964ryc}.

We first focus on the quark-antiquark annihilation channel in~\eqref{eq:qqbgaaprocess}, and describe details specific to the crossed channels~(\ref{eq:qgqaaprocess},\ref{eq:gqbaaprocess}) afterwards.
We express the scattering amplitude as
\begin{equation}
\label{eq:amplitudeqqb}
\hspace{-2mm} A_{ij}^a = i\, (4 \pi \alpha) Q_q^2 \sqrt{4 \pi \alpha_s} \,\mathbf{T}_{ij}^a \, \mathcal{A}\,.
\end{equation}
Here, $\alpha$ is the fine structure constant, $\alpha_s$ is the strong coupling constant, $Q_q$ is the quark charge in units of the electron charge,
$i,j$ are the color indices of the quark-antiquark pair, $a$ is the gluon color index and $\mathbf{T}_{ij}^a$ are the SU(3) color generators in the fundamental representation. 

We extract the helicity amplitudes for this process
following the approach suggested in~\cite{Peraro:2019cjj,Peraro:2020sfm}.
Specifically, we work in conventional dimensional regularization, but construct suitable projectors to calculate only the physical helicity amplitudes which are present in the 't Hooft-Veltman scheme.
This is achieved by decomposing the amplitude into Lorentz structures which are independent in four dimensions, thereby completely avoiding the introduction of evanescent form factors.
We express the color stripped amplitude as 
\begin{equation}
\label{eq:ampcolstrip}
\mathcal{A} =  \mathcal{A}^{\mu \nu \rho} \epsilon^*_{\mu}(p_3) \epsilon^*_{\nu}(p_4) \epsilon^*_{\rho}(p_5)\,,
\end{equation}
where $\epsilon^*_\mu(p_3)$ is the polarization vector of the gluon, and $\epsilon^*(p_j)$ with $j=4,5$ are the polarization vectors of the photons.
We impose transversality for the polarization vectors, $\epsilon_i \cdot p_i = 0$ for $i=3,4,5$, and make the cyclic choice 
$\epsilon_3 \cdot p_4 = \epsilon_4 \cdot p_5 = \epsilon_5 \cdot p_1 = 0$ to fix the gauge.
We obtain
\begin{align}
\mathcal{A} = \sum_{j=1}^{16} F_j\, \mathcal{T}_j\,, \label{eq:formfactors}
\end{align}
with the 16 Lorentz structures
\begin{align}
\mathcal{T}_{1\leq j \leq 8}  &= \bar{u}(p_2)\slashed{p}_3 u(p_1) t_j^{\mu\nu\rho}
\epsilon^*_{\mu}(p_3) \epsilon^*_{\nu}(p_4) \epsilon^*_{\rho}(p_5),
\nonumber \\
\mathcal{T}_{9\leq j \leq 16} &= \bar{u}(p_2)\slashed{p}_4 u(p_1) t_{j-8}^{\mu\nu\rho}
\epsilon^*_{\mu}(p_3) \epsilon^*_{\nu}(p_4) \epsilon^*_{\rho}(p_5),
\label{eq:Tj}
\end{align}
where the tensors $t_j^{\mu\nu\rho}$ are given by 
\begin{align}
t_1^{\mu\nu\rho} & = p_{1}^{\mu} p_{1}^{\nu} p_{2}^{\rho},
&& t_2^{\mu\nu\rho} = p_{1}^{\mu} p_{1}^{\nu} p_{3}^{\rho},\nonumber \\
t_3^{\mu\nu\rho} & = p_{1}^{\mu} p_{2}^{\nu} p_{2}^{\rho},
&& t_4^{\mu\nu\rho} = p_{1}^{\mu} p_{2}^{\nu} p_{3}^{\rho},\nonumber \\
t_5^{\mu\nu\rho} & = p_{2}^{\mu} p_{1}^{\nu}  p_{2}^{\rho},
&& t_6^{\mu\nu\rho} = p_{2}^{\mu} p_{1}^{\nu}  p_{3}^{\rho},\nonumber \\
t_7^{\mu\nu\rho} & = p_{2}^{\mu} p_{2}^{\nu} p_{2}^{\rho},
&& t_8^{\mu\nu\rho} = p_{2}^{\mu} p_{2}^{\nu} p_{3}^{\rho}\,.
\end{align}
We stress that the number of  independent Lorentz structures matches the number of helicity configurations for this process.

Each form factor $F_j$ in Eq.~\eqref{eq:formfactors} can then be computed by defining a
projector
\begin{equation}
\mathcal{P}_j = \sum_{k=1}^{16} c_{k}^j \mathcal{T}_k^\dagger\, \label{eq:projectors}
\end{equation}
with $c_k^j$ being rational functions of the Mandelstam invariants 
 such that
\begin{equation}
F_j = \sum_{\text{pol}} \mathcal{P}_j \mathcal{A}\,,
\end{equation}
where the polarization sums read
\begin{align}
    \sum_{\text{pol}} u(p)\bar{u}(p) &= \slashed{p},\\
    \sum_{\text{pol}} \epsilon_\mu(p)\epsilon^\ast_\nu(p) &= -g_{\mu \nu} + \frac{p_\mu q_\nu + p_\nu q_\mu }{p \cdot q},
\end{align}
with $q=p_4,p_5,p_6$ for $p=p_3,p_4,p_5$, respectively, and $g_{\mu\nu}$ is $d$ dimensional. We 
stress that even if, by construction, the algebra to derive the projectors 
is performed in $d$ space-time dimensions, the coefficients $c_k^j$ 
in Eq.~\eqref{eq:projectors} do not depend on $d$; see~\cite{Peraro:2019cjj,Peraro:2020sfm}
for details.

It is straightforward to compute helicity amplitudes from Eq.~\eqref{eq:formfactors} in terms of the form factors $F_j$ by evaluating the Lorentz structures of Eq.~\eqref{eq:Tj} for specific helicities of the external particles in four dimensions.
We denote the dependence of the color-stripped amplitude~\eqref{eq:ampcolstrip} on the external helicities by
\begin{equation}
\mathcal{A}(\h)\quad\text{with~} \h = \{\lambda_q, \lambda_3,\lambda_4,\lambda_5\},
\end{equation}
where $\lambda_{q} = L,R$ is the helicity of the quark line, and $\lambda_j = \pm$ for $j=3,4,5$ are the helicities of the gluon and the two photons, respectively.
For each of the partonic sub-channels that we consider, 
there are 3 independent helicity amplitudes, 
from which all the remaining ones can be obtained using parity and charge conjugation transformations 
and permutations of the external photons. 
We choose the following as independent configurations,
\begin{align}
\label{eq:refhelconf}
\h_A &= \{L,+,+,+ \}, \nonumber \\
\h_B &= \{L,-,+,+ \}, \nonumber \\
\h_C &= \{L,-,-,+ \}.
\end{align}
Note that $\mathcal{A}(\h_A)$ is zero at tree level.

We find it convenient to write each helicity amplitude by factoring out 
a helicity-dependent combination of spinor products which carries the spinor weight of the amplitude. 
In practice, this can be achieved by factoring out the tree-level amplitude in case it is non-zero,
and an arbitrary combination of spinor products with the correct spinor weight otherwise.
It is particularly convenient to carry out this factorization using the spinor-helicity formalism. 
Explicitly, we write for left-handed spinors $\bar{u}_L(p_2) = \langle 2 |$ and $u_L(p_1) = | 1 ]$
and for the gauge boson $j$ of momentum $p_j$
\begin{equation}
\epsilon^\mu_{j,-}(q_j) = \frac{\langle q_j | \gamma^\mu | j ] }{ \sqrt{2} \langle q_j j \rangle}\,, \quad
\epsilon^\mu_{j,+}(q_j) = \frac{\langle j | \gamma^\mu | q_j ] }{ \sqrt{2} [ j q_j  ]}\,,
\end{equation}
where $q_j$ is the gauge fixing momentum.
With these, we define spinor-free amplitudes
$\mathcal{B}(\h)$ according to
\begin{equation}
\mathcal{A}(\h)= \Phi(\h) \mathcal{B}(\h)\,,
\end{equation}
where for the spinor functions  $ \Phi(\h)$ we choose~\cite{DelDuca:2003uz,Abreu:2021fuk}
\begin{align}
\Phi(\h_A) &= 2 \sqrt{2} \frac{[31] \langle12 \rangle^3  \langle 13\rangle}{ \langle 14\rangle^2  \langle 15\rangle^2  \langle 23\rangle^2}\,, \nonumber \\
\Phi(\h_B) &=  2 \sqrt{2} \frac{ \langle 12 \rangle  \langle 23\rangle^2 }{ \langle 14\rangle \langle42 \rangle \langle25 \rangle \langle51 \rangle}\,, \nonumber \\
\Phi(\h_C) &=    2 \sqrt{2}\frac{ [51]^2 [12]}{[14][42][23][31]}\,.
\end{align}
The corresponding phases 
for the $qg$ and $g\bar{q}$ initiated processes can be obtained by exchanging
$2\leftrightarrow3$ and $1\leftrightarrow3$ respectively.

In order to calculate the spinor-free amplitudes $\mathcal{B}(\h)$ it is useful to
decompose them into parity even and odd contributions,
\begin{equation}
\mathcal{B}(\h)= \mathcal{B}^E(\h) + \overline{\epsilon}_5\,  \mathcal{B}^O(\h)\,,
\end{equation}
where $\overline{\epsilon}_5 = \epsilon_5/{s^2_{12}}$ is dimensionless.
We stress here
that $\overline{\epsilon}_5$ changes sign under parity transformations or 
odd permutations of the external momenta; see Eq.~\eqref{eq:epsfive}.
The $\mathcal{B}^P(\h)$ terms, with $P=E,O$, can in turn be expressed as 
linear combinations of the form factors
in Eq.~\eqref{eq:formfactors} with coefficients that depend only on the $s_{ij}$~\cite{Peraro:2019cjj,Peraro:2020sfm}.
From the projection operators in Eq.~\eqref{eq:projectors}
one can then derive six independent \textsl{helicity projector operators}, decomposed in terms of the complex conjugates of the tensors in Eq.~\eqref{eq:Tj}, 
and which directly project onto the 
$\mathcal{B}^P(\h)$ defined above.
The explicit form of these projectors 
can be obtained from the authors upon request.

The QCD corrections to the helicity amplitudes can be computed
by expanding in the bare strong coupling constant $\alpha^b_s$,
\begin{align}
\mathcal{B}^P(\h) = 
\sum_{k=0}^2 \left( \frac{\alpha^b_s}{2 \pi}\right)^k \mathcal{B}^{P,(k)}(\h)
+ \mathcal{O}((\alpha^b_s)^3)\,.
\end{align}
By construction, the first order reads
\begin{align}
&\mathcal{B}^{E,(0)}(\h_A) = 0 \,, \quad \mathcal{B}^{O,(0)}(\h_A)= 0 \nonumber \\
&\mathcal{B}^{E,(0)}(\h_B) = 1 \,, \quad \mathcal{B}^{O,(0)}(\h_B) = 0  \nonumber \\
&\mathcal{B}^{E,(0)}(\h_C) = 1 \,, \quad \mathcal{B}^{O,(0)}(\h_C) = 0\,. 
\end{align}
We calculate the one- and two-loop corrections $\mathcal{B}^{P,(k)}$, $k=1,2$, as follows.
First, we generate all relevant Feynman diagrams using \texttt{Qgraf}~\cite{Nogueira:1991ex}.
We then contract them with the helicity projectors described above 
and express the amplitudes in terms
of scalar Feynman integrals.
All the algebra required at this level has been carried out using \texttt{FORM}~\cite{Vermaseren:2000nd}.
The two-loop corrections, in particular, can be mapped to a large number of scalar two-loop integrals drawn 
from two different integral families (see e.g.~\cite{Gehrmann:2018yef,Agarwal:2021grm})
\begin{equation}
\mathcal{I}^{\rm fam}_{n_1,...,n_{11}} = 
e^{2 \epsilon \gamma_E }\, \int \prod_{i=1}^2 \left(\frac{\rd^d k_i}{i \pi^{d/2}}\right) \frac{1}{D_1^{n_1} ... D_{11}^{n_{11}} }\,,
\end{equation}
where $d=4-2\epsilon$ is the space-time dimension and $\gamma_E \sim 0.5772$ is the Euler-Mascheroni constant, and the $D_i$ are the loop propagators. 
We provide the definition of the two integral families ${\rm fam} = \{A,B\}$ in the supplemental material.
In order to describe all integrals contributing to the amplitudes, we also require crossed families, which are obtained from the two reference families described above by permutations of external momenta.

As it is well known, the scalar integrals fulfil linear relations and can be reduced to a set of  master integrals
using symmetry relations and integration-by-parts identities~\cite{Tkachov:1981wb,Chetyrkin:1981qh}.
Laporta's algorithm~\cite{Laporta:2001dd} maps the problem to the 
row reduction of a large matrix.
While straightforward in principle,
this step can become computationally very challenging and, until now, has remained a major bottleneck for the calculation of the non-planar two-loop corrections to $2 \to 3$ massless scattering amplitudes.\footnote{Very recently, the reduction of the most complicated non-planar rank five integrals has been achieved independently in~\cite{Bendle:2021ueg}.} 
We succeed in reducing all planar and non-planar two-loop Feynman integrals as follows.
First, we use \texttt{Reduze\;2}~\cite{vonManteuffel:2012np} to identify shift, symmetry and crossing relations of the scalar integrals.
The actual integration-by-parts reduction of the remaining integrals is performed using a private implementation of the Laporta algorithm,
\texttt{Finred}, augmented by the use
of finite-field arithmetics \cite{Wang:1982proof,vonManteuffel:2014ixa,Peraro:2016wsq}, syzygy techniques \cite{Gluza:2010ws,Ita:2015tya,Larsen:2015ped,Boehm:2017wjc,Agarwal:2020dye} and denominator guessing \cite{Abreu:2018zmy,Heller:2021qkz}.
Despite expressing the integrals directly in terms of the canonical basis defined in~\cite{Chicherin:2020oor}, the reduction identities for the non-planar integrals are quite cumbersome if the rational coefficients are 
represented in a common-denominator form, for example, with \texttt{Fermat}~\cite{fermat}.
As it has already been observed in previous work~\cite{Abreu:2019odu,Boehm:2020ijp,Chawdhry:2020for,Heller:2021qkz,Agarwal:2021grm}, substantial simplifications of these identities can be achieved
with a multivariate decomposition into partial fractions of the relevant rational
coefficients.\footnote{As exemplification, the most complicated reduction identities 
in our calculation are for rank-five non-planar double-pentagon integrals and have sizes of order $1\,$GB each.
Multivariate partial fractioning reduces their size by factors of order 100.}
In practice, we find it most efficient to first perform the reduction for a minimal subset of integrals in the uncrossed families $\{A,B\}$, 
simplify them using partial fractioning, and finally cross them.
For the partial fraction decomposition we employ \texttt{MultivariateApart}~\cite{Heller:2021qkz}, where we use \texttt{Singular}~\cite{DGPS} as a backend for the polynomial reductions.
In this way, we produce all required identities for the reduction of the amplitude in a compact representation.
After inserting the reduction identities in the amplitude we perform another partial fraction decomposition.
We note that in addition to the denominator factors relevant for the leading color interferences \cite{Agarwal:2021grm}, the full color helicity amplitudes contain also the Gram determinant in  Eq.~\eqref{eq:Gram} as a denominator factor.
Finally, we express our results in terms of the pentagon functions defined in~\cite{Chicherin:2020oor}. 

The bare helicity amplitudes contain poles in the dimensional regulator $\epsilon$ 
both of ultraviolet (UV) and infrared (IR) origin, up to order $\epsilon^{-4}$.
We remove UV singularities by expressing our
result in terms of the $\overline{ \rm MS}$ renormalized strong coupling $\alpha_s(\mu)$: 
\begin{align}
\mathcal{B}^P(\h) = 
\sum_{k=0}^2 \left( \frac{\alpha_s(\mu)}{2 \pi}\right)^k \mathcal{\overline{B}}^{P,(k)}(\h)
+ \mathcal{O}(\alpha_s^3)\,,
\end{align}
where the relation between renormalized and bare coupling is given by
\begin{equation}
S_{\epsilon}\alpha_{s} = \mu^{2\epsilon} \alpha_s(\mu) Z[{\alpha_s(\mu)}],
\end{equation}
with $S_{\epsilon} = (4\pi)^{-\epsilon}e^{-\gamma_E \epsilon}$ and
\begin{equation}
Z[\alpha] = 1-\frac{\beta_0}{\epsilon}\left(\frac{\alpha_s}{2\pi}\right) + 
\left(\frac{\beta_0^2}{\epsilon^2}-\frac{\beta_1}{2\epsilon}\right)
\left(\frac{\alpha_s}{2\pi}\right)^2+ \mathcal O(\alpha_s^3)\,.
\end{equation}
The coefficients of the beta function read
$$\beta_0 = \frac{11}{6}C_A - \frac{2}{3} T_r n_f\,, 
\;\;
\beta_1 = \frac{17}{6}C_A^2 - \left(\frac{5}{3} C_A  - C_F \right)T_r n_f\,,$$
where we use $T_r=1/2$.
The UV-renormalized amplitudes $\mathcal{\overline{B}}^{P,(k)}(\h)$ still contain divergences of IR origin.
Following~\cite{Catani:1998bh}, their structure can be universally expressed in terms
of the lower-order scattering amplitudes as
\begin{align} \label{eq:finremI}
&\mathcal{\overline{B}}^{P,(1)}(\h) = \mathcal{I}_1\, \mathcal{\overline{B}}^{P,(0)}(\h) + \mathcal{R}^{P,(1)}(\h) \, \\
&\mathcal{\overline{B}}^{P,(2)}(\h) = \mathcal{I}_2\, \mathcal{\overline{B}}^{P,(0)}(\h)  +
\mathcal{I}_1\, \mathcal{\overline{B}}^{P,(1)}(\h)  + \mathcal{R}^{P,(2)}(\h)\,, \nonumber
\end{align}
where the $\mathcal{I}_k$ are universal and depend only on the loop order and 
on the species of the colored partons in the initial and final state of the scattering process,
and the $\mathcal{R}^{P,(k)}(\h)$ are the so-called finite remainders.
The explicit expressions of the operators $\mathcal{I}_k$ with $k=1,2$ are provided
in the supplemental material.

It is convenient to organize the one- and two-loop corrections to the finite remainders $\mathcal{R}^{P,(k)}(\h)$ into separately gauge-invariant color structures
\begin{align}
\label{eq:finiteremainders}
&\mathcal{R}^{P,(1)}(\h) = \sum_{i=1}^4 \tilde{b}_i\,   \mathcal{R}^{P,(1)}_i(\h) \nonumber \\
&\mathcal{R}^{P,(2)}(\h) = \sum_{i=1}^{10} \tilde{c}_i\, \mathcal{R}^{P,(2)}_i(\h) \,.
\end{align}
At one-loop we choose the following basis,
\begin{align}
\tilde{b}_1 = N\,, \quad \tilde{b}_2 = N^{-1}\,, \quad \tilde{b}_3 = n_f^{\gamma \gamma}, \quad \tilde{b}_4 = n_f,
\end{align}
whereas at two-loops we have
\begin{align} \label{eq:cf2l}
& \tilde{c}_1 = N^2,   && \tilde{c}_2 = 1,          && \tilde{c}_3 =N^{-2}, \nonumber \\
& \tilde{c}_4 = N n_f, && \tilde{c}_5 = N^{-1} n_f, && \tilde{c}_6 =n^{\gamma\gamma}_f n_f, \nonumber\\
& \tilde{c}_7 = N n^{\gamma\gamma}_f,       && \tilde{c}_8 = N^{-1} n^{\gamma\gamma}_f,  
&& \tilde{c}_9 = n^\gamma_f\left(N - 4N^{-1} \right) \nonumber\\
& \tilde{c}_{10} = n^2_f,
\end{align}
where $N$ is the number of colors, 
$n_f$ is the number of massless quarks in closed fermionic loops,
and $\nfvo$ and $\nfvt$ are defined as
\begin{equation}
n_f^{\gamma} = \frac{1}{Q_q} \sum^{n_f}_{i=1} Q_i, \quad\;
n_f^{\gamma \gamma} = \frac{1}{Q^2_q}\sum^{n_f}_{i=1} Q^2_i\,.
\end{equation}
\begin{table}[t]
\begin{center}
\begin{tabular}{| l | r | r |}
\hline
& \multicolumn{1}{|c|}{$u\bar{u}\to g \gamma\gamma$}
& \multicolumn{1}{|c|}{$u g\to u \gamma\gamma$}  \\
\hline
$\calR^{\!(1)}\!(\h_{A})$   & $ 0.08637873 + 0.6505825\,i$ & $-0.05575262 + 1.282163\,i$ \\
$\calR^{\!(1)}\!(\h_{B})$   & $   4.812087 + 0.8811173\,i$ & $  -5.332701 - 6.518506\,i$\\
$\calR^{\!(1)}\!(\h_{C})$   & $ 0.05297897 -  4.432186\,i$ & $  -2.497722 - 22.42864\,i$\\
\hline
$\calR^{\!(2)}\!(\h_{A})$   & $-2.385158 + 18.22971\,i$    & $-28.12588 + 26.67761\,i$\\
$\calR^{\!(2)}_{\LC}\!(\h_{A})$   & $0.4123777 + 22.64313\,i$    & $-1.450073 + 7.396238\,i$\\
$\calR^{\!(2)}\!(\h_{B})$   & $ 115.9528 + 18.71704\,i$    & $ 17.16557 - 102.3377\,i$\\
$\calR^{\!(2)}_{\LC}\!(\h_{B})$   & $ 144.2892 - 3.600533\,i$    & $ 33.14649 - 134.9655\,i$\\
$\calR^{\!(2)}\!(\h_{C})$   & $-36.87656 - 153.3540\,i$    & $-26.92189 - 508.2138\,i$\\
$\calR^{\!(2)}_{\LC}\!(\h_{C})$   & $-55.57522 - 190.2039\,i$    & $ 76.13565 - 214.1456\,i$\\
\hline
\end{tabular}
\caption{
One- and two-loop finite remainders for diphoton plus parton production processes.
Samples are shown for all independent helicity configurations $\h$ \eqref{eq:refhelconf} 
in the kinematic point \eqref{eq:benchmarkpoint},
either including full color, $\mathcal{R}^{(i)}(\h)$, $i=1,2$, 
\eqref{eq:finiteremainderEO}, or only two-loop leading color, $\mathcal{R}_\text{LC}^{(2)}(\h)$ \eqref{eq:finiteremainderEOlc} contributions.
}
\label{tab:benchmarkresults}
\end{center}
\end{table}
The finite remainders $\mathcal{R}^{P,(k)}_i(\h)$ are formally expressed as linear combinations
of the pentagon functions defined in~\cite{Chicherin:2020oor}, whose coefficients 
are rational functions of the kinematic invariants $s_{ij}$. After finding a minimal set of
such rational functions, we perform an optimized multivariate partial fraction decomposition 
that allows us to avoid not only spurious denominators, but also unnecessarily high denominator powers throughout. This two-step procedure, detailed in \cite{Agarwal:2021grm}, renders our
final expressions very compact and optimizes them for fast and stable numerical evaluations.
Notably, for the most complicated color factor which involves non-planar contributions, $c_2$ in Eq.~\eqref{eq:cf2l}, the largest helicity coefficient amounts to only 4.5\,MB in size.

The discussion presented so far focused on the $q\bar{q}$ partonic channel 
in Eq.~\eqref{eq:qqbgaaprocess}, but the formal expressions of Eq.~\eqref{eq:finiteremainders}
hold for the crossed channels in Eqs.~(\ref{eq:qgqaaprocess},\ref{eq:gqbaaprocess}) as well.
In fact, results for these processes can be readily extracted from the ones in the $q\bar{q}$ channel
without repeating any of the heavy parts of the calculation.
In practice, we obtain the helicity amplitudes
for the $g\bar{q}$ and $qg$ channels from the $q\bar{q}$ amplitudes by applying the permutations $1\leftrightarrow3$ and $2\leftrightarrow3$, respectively.
While this operation is trivial for rational functions of the $s_{ij}$ and
$\overline{\epsilon}_5$, this is not the case for the pentagon functions.
In order to express the crossed pentagon functions in terms of the uncrossed ones,
we proceed as follows.
First, we express the two-loop master integrals and crossings thereof in terms of pentagon functions.
Second, we employ the fact that the full set of master integrals is linearly mapped onto itself by
any crossing of the external legs.
By combining and reducing these systems of equations, we obtain reduction identities for pentagon functions with $1\leftrightarrow3$ and $2\leftrightarrow3$ permuted, respectively.
These solutions provide us with sufficient information to successfully cross the helicity amplitudes and express them in terms of the original pentagon functions.

We perform various checks on our results.
First, we verify that our tree-level and one-loop helicity amplitudes agree with the \texttt{OpenLoops\;2} program~\cite{Buccioni:2019sur}.
Next, we check that the poles of our two-loop amplitudes follow the pattern predicted by Catani 
and we compare the leading color part of our two-loop helicity amplitudes with the results of~\cite{Chawdhry:2021mkw} and find agreement.
Finally, we perform a separate calculation of the interference of the two-loop and tree-level amplitudes summed over polarizations, both for the $q\bar{q}$ and for the $qg$ channels.
In this second calculation no projectors are used, and the $qg$ channel is obtained by crossing the relevant amplitudes prior to reduction to master integrals instead of crossing the pentagon functions.
After subtracting the UV and IR poles, we find perfect agreement for the four-dimensional finite remainder of all color factors between 
the two calculations, which provides a very strong check of  
our helicity amplitudes.

In Tab.~\ref{tab:benchmarkresults} we present benchmark results for the full-color one- and two-loop finite remainders
\begin{align}
\label{eq:finiteremainderEO}
\mathcal{R}^{(i)}(\h) &= \mathcal{R}^{E,(i)}(\h) + \bar{\epsilon}_5\mathcal{R}^{O,(i)}(\h)\quad (i=1,2),\\
\intertext{and the leading-color two-loop finite remainder}
\label{eq:finiteremainderEOlc}
\mathcal{R}_{\text{LC}}^{(2)}(\h) &= \tilde{c}_1 \lb \mathcal{R}_1^{E,(2)}(\h) + \bar{\epsilon}_5\mathcal{R}_1^{O,(2)}(\h) \rb,
\end{align}
for the three independent helicity configurations,
where we choose $q=u$ (up-quark), $n_f=5$, the kinematic point
\begin{align} 
  s_{12}&=157, \quad   &s_{23}&=-43, \quad &s_{34}&=83, \nonumber \\
  s_{45}&=61,  \quad   &s_{15}&=-37, \quad &\mu^2 &= 100\,,
  \label{eq:benchmarkpoint}
\end{align}
and use the \texttt{PentagonMI} package~\cite{Chicherin:2020oor} for the numerical evaluation of the pentagon functions.
At \url{https://gitlab.msu.edu/vmante/aajamp-symb}, we provide our complete analytical results for the one- and two-loop finite remainders for all helicity configurations and color factors, separated into parity even and odd contributions.

To conclude, in this Letter we have presented the first calculation of the
full-color two-loop QCD corrections to all helicity
amplitudes of a $2 \to 3$ scattering process.
Our analytical calculation relied on recently developed techniques to handle helicity amplitudes
of multileg processes and to perform the required reductions to master integrals.
With the results for diphoton plus jet production published here, all amplitudes required for the calculation of N$^3$LO observables for diphoton production at the LHC are now available.

{\bf Acknowledgments}
We would like to thank Rene Poncelet for facilitating the comparison of the
leading color part against the results in~\cite{Chawdhry:2021mkw}.
We further thank Fabrizio Caola and Tiziano Peraro for useful comments on the manuscript.
BA and AvM are supported in part by the National Science Foundation through Grant 2013859.
The research of FB is supported by the ERC Starting Grant 804394 HipQCD.
LT is supported by the Royal Society through Grant URF/R1/191125.

\bibliography{qqaagfull}

\newpage

\onecolumngrid

\section*{Supplemental material}

\makeatletter
\renewcommand\@biblabel[1]{[#1S]}
\makeatother

\subsection{Integral families}

In the following table, we provide the definition of the two integral families that we use to
perform the calculation of the scattering amplitudes for $q\bar{q} \to g \gamma \gamma$, $qg \to q\gamma \gamma$ and  $g\bar{q}\to \bar{q}\gamma \gamma$:
\begin{center}
\begin{tabular}{| c | l | l|}
\hline
Prop.\ den.\ & Family A & Family B  \\
\hline
\hline
$D_1$ & $ k_1^2$ &                                        $k_1^2$    \\
$D_2$ & $(k_1+p_1)^2$ &                              $(k_1-p_1)^2$   \\
$D_3$ & $(k_1+p_1+p_2)^2$ &                      $(k_1-p_1-p_2)^2$   \\
$D_4$ & $(k_1+p_1+p_2+p_3)^2$ &              $(k_1-p_1-p_2-p_3)^2$   \\
$D_5$ & $k_2^2 $ &                                        $k_2^2 $   \\
$D_6$ & $(k_2+p_1+p_2+p_3)^2$ &               $(k_2-p_1-p_2-p_3-p_4)^2$   \\
$D_7$ & $(k_2+p_1+p_2+p_3+p_4)^2$ &       $(k_1-k_2)^2$   \\
$D_8$ & $(k_1 - k_2)^2$ &                               $(k_1-k_2+p_4)^2$  \\
$D_9$ & $(k_1+p_1+p_2+p_3+p_4)^2$ &       $(k_2-p_1)^2$   \\
$D_{10}$ & $(k_2 + p_1 )^2$ &                        $(k_2-p_1-p_2)^2$   \\
$D_{11}$ & $(k_2 +p_1 +p_2)^2$ &                 $(k_2-p_1-p_2-p_3)^2$   \\
\hline 
\end{tabular}
\end{center}
Note that family B alone is sufficient to describe all non-planar topologies.

\subsection{Infrared structure}
In this Appendix, we provide the explicit form of Catani's operators, which we employ to
define the one- and two-loop finite remainders of the helicity amplitudes; see Eqs.~(\ref{eq:finremI},\ref{eq:finiteremainders}).
For the $q\bar{q} \to g \gamma \gamma$ channel we have
\begin{align}
    \mathcal{I}_1 = \frac{\mathrm{e}^{\eps\gamma_{E}}}{2\Gamma(1-\eps)}
    \bigg[ &\lb C_A - 2 C_F\rb \lb \frac{1}{\eps^2} + \frac{3}{2\eps} \rb
    \lb-\frac{\mu^2}{s_{12}}\rb^\eps
    - \lb C_A \lb \frac{1}{\eps^2} + \frac{3}{4\eps} \rb + \frac{\gamma_g}{2\eps} \rb
    \lb \lb-\frac{\mu^2}{s_{13}}\rb^\eps + \lb-\frac{\mu^2}{s_{23}}\rb^\eps \rb\bigg],
\label{eq:CataniIone}
\end{align}
with $\gamma_g = \beta_0$ and 
\begin{align}
    \mathcal{I}_2 =  -\frac{1}{2} \mathcal{I}_1\lb \mathcal{I}_1 + 2\frac{\beta_0}{\eps} \rb 
+    \mathrm{e}^{-\eps\gamma_{E}}
    \frac{\Gamma(1-2\eps)}{\Gamma(1-\eps)}\lb\frac{\beta_0}{\eps}+K\rb \mathcal{I}_1 +
    \frac{\mathrm{e}^{-\eps\gamma_{E}}}{4\eps\Gamma(1-\eps)} H_2,
\end{align}
where $K$ is universal,
\begin{equation}
    K = \lb \frac{67}{18} - \frac{\pi^2}{6}\rb C_A - \frac{10}{9} T_r n_f,
\end{equation}
and $H_2$ depends on the renormalization procedure and on the process at hand.
In our case we have
\begin{equation}
H_2 = 2 H_{2,q} + H_{2,g},
\end{equation}
with $H_{2,q}$ and $H_{2,g}$ defined as
\begin{align}
    H_{2,g} &= \lb \frac{\zeta_3}{2} + \frac{5}{12} + \frac{11\pi^2}{144} \rb C^2_A +
    \frac{20}{27}T_r^2 n_f^2 - \lb \frac{\pi^2}{36} + \frac{58}{27}\rb C_A T_r n_f + C_F T_r n_f, \\
    H_{2,q} &= \lb -\frac{3}{8} + \frac{\pi^2}{2} - 6 \zeta_3 \rb C^2_F + 
    \lb \frac{245}{216} - \frac{23\pi^2}{48} + \frac{13\zeta_3}{2} \rb C_A C_F
    + \lb\frac{\pi^2}{12} - \frac{25}{54}\rb C_F T_r n_f\, .
\end{align}
In Eq.~\eqref{eq:CataniIone} the $s_{12} = s_{12} + i 0^+$ prescription is understood, and $s_{13}$ is
defined as $s_{13} = (p_1 - p_3)^2$, so that $s_{13} < 0$.
In order to obtain the corresponding expressions for the $qg$ and $g \bar{q}$ channels,
it is sufficient to perform the permutations $2\leftrightarrow 3$ and $1\leftrightarrow 3$ in Eq.~\eqref{eq:CataniIone}, respectively.

\subsection{Remaining helicity amplitudes}

In this Appendix, we show explicitly how to obtain all the remaining helicity amplitudes
from the reference amplitudes in Eq.~\eqref{eq:refhelconf}. The same discussion extends to 
the finite remainders in Eq.~\eqref{eq:finiteremainderEO} as well.

For the $q\bar{q}\to g \gamma\gamma$ channel we have
\begin{align*}
  \calA(\left\lbrace R,-,-,- \right\rbrace) &= \Phi^{*}(\h_{A}) \times \mathrm{P}\left[\calB(\h_{A})\right], \\
  \calA(\left\lbrace R,+,+,+ \right\rbrace) &= \Phi_{\mathrm{x12}}(\h_{A}) \times \mathrm{P_{x12}}\left[\calB(\h_{A})\right],\\
  \calA(\left\lbrace L,-,-,- \right\rbrace) &= \Phi^{*}_{\mathrm{x12}}(\h_{A}) \times \mathrm{P}\left[\mathrm{P_{x12}}\left[\calB(\h_{A})\right]\right],\\
  \calA(\left\lbrace R,+,-,- \right\rbrace) &= \Phi^{*}(\h_{B}) \times \mathrm{P}\left[\calB(\h_{B})\right], \\
  \calA(\left\lbrace R,+,+,- \right\rbrace) &= \Phi^{*}(\h_{C}) \times \mathrm{P}\left[\calB(\h_{C})\right], \\
  \calA(\left\lbrace R,-,+,+ \right\rbrace) &= \Phi_{\mathrm{x12}}(\h_{B}) \times \mathrm{P_{x12}}\left[\calB(\h_{B})\right], \\
  \calA(\left\lbrace R,-,-,+ \right\rbrace) &= \Phi_{\mathrm{x12}}(\h_{C}) \times \mathrm{P_{x12}}\left[\calB(\h_{C})\right], \\
  \calA(\left\lbrace L,+,-,- \right\rbrace) &= \Phi^{*}_{\mathrm{x12}}(\h_{B}) \times \mathrm{P}\left[\mathrm{P_{x12}}\left[\calB(\h_{B})\right]\right],\\
  \calA(\left\lbrace L,+,+,- \right\rbrace) &= \Phi^{*}_{\mathrm{x12}}(\h_{C}) \times \mathrm{P}\left[\mathrm{P_{x12}}\left[\calB(\h_{C})\right]\right],\\
  \calA(\left\lbrace L,-,+,- \right\rbrace) &= \Phi_{\mathrm{x45}}(\h_{C}) \times \mathrm{P_{x45}}\left[\calB(\h_{C})\right], \\
  \calA(\left\lbrace R,-,+,- \right\rbrace) &= \Phi_{\mathrm{x12x45}}(\h_{C}) \times \mathrm{P_{x12x45}}\left[\calB(\h_{C})\right], \\
  \calA(\left\lbrace R,+,-,+ \right\rbrace) &= \Phi^{*}_{\mathrm{x45}}(\h_{C}) \times \mathrm{P}\left[\mathrm{P_{x45}}\left[\calB(\h_{C})\right]\right], \\
  \calA(\left\lbrace L,+,-,+ \right\rbrace) &= \Phi^{*}_{\mathrm{x12x45}}(\h_{C}) \times \mathrm{P}\left[\mathrm{P_{x12x45}}\left[\calB(\h_{C})\right]\right],
\end{align*}
where the operators $\mathrm{P}$, $\mathrm{P_{x12}}$, $\mathrm{P}_{x45}$ and $\mathrm{P_{x12x45}}$ implement
the parity transformation, the odd permutations $(1 \leftrightarrow 2)$, $(4 \leftrightarrow 5)$
and the even permutation $(1 \leftrightarrow 2, 4 \leftrightarrow 5)$ of external momenta, respectively, \begin{align*}
    \mathrm{P}\left[\calB(\h)\right] & = \calB^{E}(\h) - \bar{\epsilon}_5 \calB^{O}(\h), \\
    \mathrm{P_{x12}}\left[\calB(\h)\right] & = \big[\calB^{E}(\h)\big]_{1\leftrightarrow 2} - \bar{\epsilon}_5 \big[\calB^{O}(\h)\big]_{1\leftrightarrow 2}, \\
    \mathrm{P_{x45}}\left[\calB(\h)\right] & = \big[\calB^{E}(\h)\big]_{4\leftrightarrow 5} - \bar{\epsilon}_5 \big[\calB^{O}(\h)\big]_{4\leftrightarrow 5}, \\
    \mathrm{P_{x12x45}}\left[\calB(\h)\right] & = \big[\calB^{E}(\h)\big]_{1\leftrightarrow 2, 4\leftrightarrow 5} + \bar{\epsilon}_5 \big[\calB^{O}(\h)\big]_{1\leftrightarrow 2, 4\leftrightarrow 5}\,.
\end{align*}
As for the spinor functions $\Phi(\h)$, the subscripts denote the permutation at hand, 
whereas the superscript $^{*}$ denotes complex conjugation, 
whose action is such that $\langle ij\rangle \leftrightarrow [ij]$.

For the $qg\to q \gamma \gamma$ and $g \bar{q}\to \bar{q} \gamma \gamma$ channels, similar relations can be established. However, to conveniently match the conventions of the \texttt{PentagonFunctions++} library~\cite{Chicherin:2020oor},
one wants to avoid the permutation of initial with final momenta.
Let us take the $qg$ channel as an example.
Considering the $\mathrm{P_{x45}}$ permutation and the parity transformation, 
one can derive five more helicity configurations. The remaining eight ones, 
where the helicity of the quarks is flipped, can be related to the $\bar{q}g$ channel, which
in turn can be derived from the $g\bar{q}$ channel via the $\mathrm{P_{x12}}$ permutation.
\end{document}